\documentclass{mem}
\usepackage{natbib}
\usepackage{txfonts}
\usepackage{balance}
\usepackage{graphicx}
\usepackage{xspace}
\usepackage[breaklinks]{hyperref}

\newcommand{\msun}{\ensuremath{M_{\odot}}\xspace}
\newcommand{\vinf}{\ensuremath{v_{\infty}}\xspace}
\newcommand{\vx}{Vela~X-1\xspace} 

\begin{document}

\title{
Vela X-1 as a laboratory  
for accretion in High-Mass X-ray Binaries
}

   \subtitle{}

\author{
P.~Kretschmar\inst{1} 
\and
S. Mart\'inez-N\'u\~nez\inst{2}
\and 
F. F{\"u}rst\inst{1}
\and
V. Grinberg\inst{3}
\and
M. Lomaeva\inst{4}
\and
I. El Mellah\inst{5}
\and
A. Manousakis\inst{6}
\and 
A.\,A.\,C. Sander\inst{7}
\and
N. Degenaar\inst{8}          
\and
J. van den Eijnden\inst{8}
}

\institute{%
 European Space Astronomy Centre (ESA/ESAC), Science Operations Department,
 E-28692, Villanueva de la Ca\~{n}ada, Madrid, Spain
\email{Peter.Kretschmar@esa.int}
\and
Instituto de F\'isica de Cantabria (CSIC-Universidad de Cantabria), E-39005, Santander, Spain 
\email{silvia.martinez.nunez@gmail.com}  
\and 
Institut f{\"u}r Astronomie und Astrophysik, Universit{\"a}t T{\"u}bingen, Sand 1, 72076 T{\"u}bingen, Germany
\and 
European Space Research and Technology Centre (ESA/ESTEC), Keplerlaan 1, 2201 AZ Noordwijk, The Netherlands
\and
Centre for Mathematical Plasma Astrophysics, Department of Mathematics, KU Leuven, 
      Celestijnenlaan 200B, 3001 Leuven, Belgium
\and
University of Sharjah, Sharjah, United Arab Emirates
\and 
Armagh Observatory and Planetarium, College Hill, Armagh, BT61 9DG, UK
\and 
Anton Pannekoek Institute for Astronomy, University of Amsterdam, Science Park 904, NL-1098 XH Amsterdam, The Netherlands      
}

\authorrunning{Kretschmar et al,}

\titlerunning{Vela X-1 as laboratory for accretion in HMXB}

\abstract{%
\vx is an eclipsing high mass X-ray binary (HMXB) consisting of a
283s accreting X-ray pulsar in a close orbit of 8.964 days around the
B0.5Ib supergiant HD77581 at a distance of just 2.4 kpc. The system is
considered a prototype of wind-accreting HMXB and it has been used as
a baseline in different theoretical or modelling studies.

We discuss the observational properties of the system and the use of
the observational data as laboratory to test recent developments in
modelling the accretion process in High-Mass X-ray Binaries 
\citep[e.g.,][]{Sander:2018,El-Mellah:2018}, which range
from detailed descriptions of the wind acceleration to modelling of
the structure of the flow of matter close to the neutron star and its
variations.
\keywords{accretion, accretion disks - stars: neutron – stars: winds – X-rays: binaries - X-rays: individual Vela X-1}
}
\maketitle{}

\begin{figure}[th]
\vspace*{-3mm}
\includegraphics[clip=true,angle=-90,width=\columnwidth]{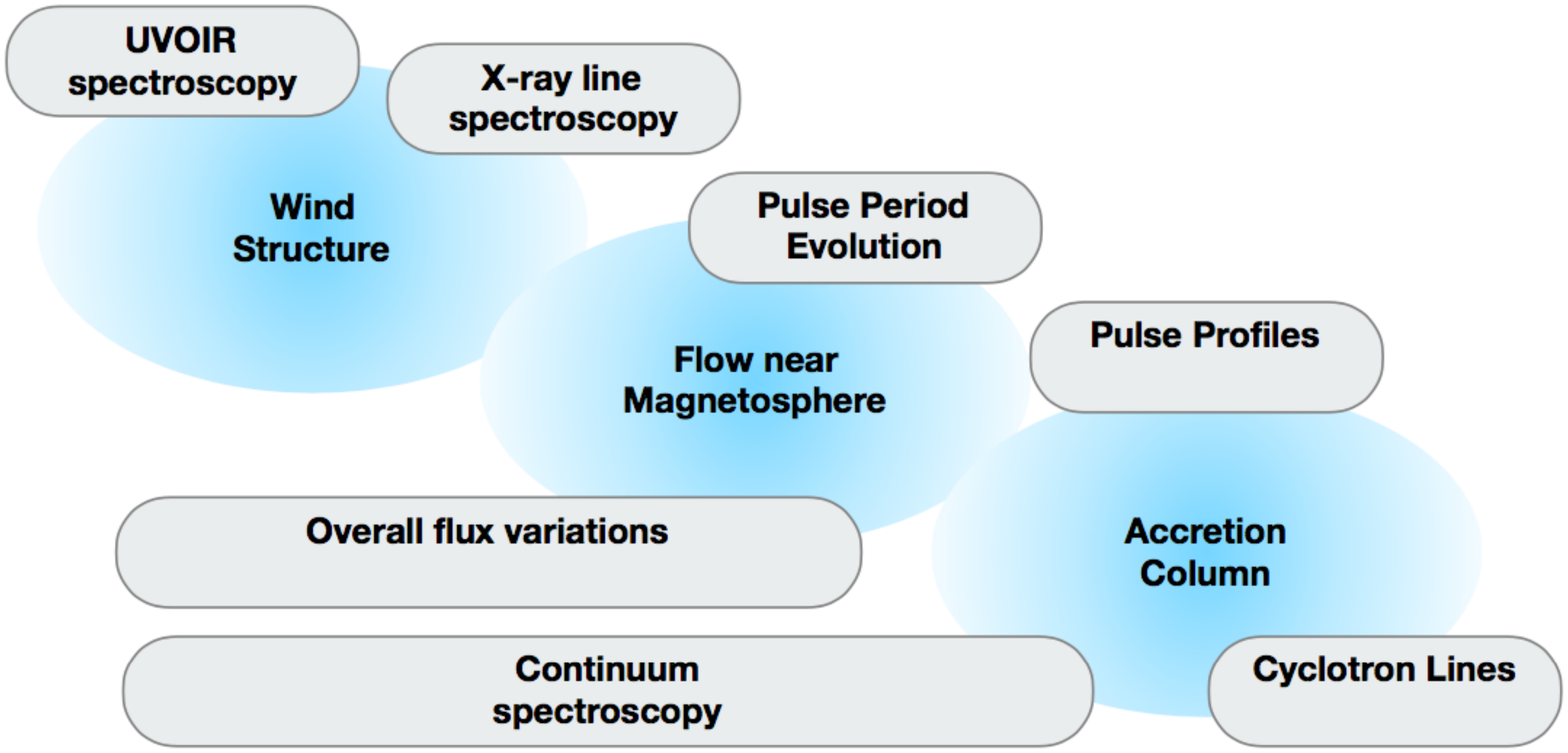}
\vspace*{-3mm}
\caption{%
Different diagnostics relate to different elements of the system.
}
\label{fig:diag}
\end{figure}
\section{Introduction}
\vx belongs to the earliest and best known X-ray binaries in the Galaxy. It was discovered by \citet{Chodil:67} in sounding rocket flights undertaken in July and September 1966. 
\citet{Forman:73} identified previously found brightness variations \citep{Ulmer:72} as caused by eclipses. 
\citet{RappaportMcClintock:75} detected that the X-ray emission was pulsed. 
In the following decades, up to the present, \vx has been observed by every major X-ray instrument. 
Table~\ref{tab:system} summarises the system parameters.
Compared to most other X-ray binaries, the essential parameters are very well known in \vx. This allows to compare modelling efforts in greater detail with the data.

\begin{table*}[ht]
\renewcommand*{\arraystretch}{1.1}
\caption{Vela X-1 system parameters}
\label{tab:system}
\begin{tabular}{lrl}
\hline
Distance           & 2.42 (2.25-–2.60) kpc     & \cite{Bailer-Jones:2018} \\
Orbital period     & $8.964357\pm 0.000029$ d   & \cite{Kreykenbohm:2008} \\
Eccentricity       & $0.0898\pm0.0012$         & \cite{Bildsten:97} \\
Inclination        & $i>79$~deg               & \cite{GimenezGarcia:2016} \\
$a \sin i$  & 113.89 lt-sec             & \cite{Bildsten:97} \\
Donor star type    & B0.5Ia                    & \cite{GimenezGarcia:2016} \\ 
Donor star mass    &  $21.5\pm 4$ \msun         & \cite{GimenezGarcia:2016} \\
Neutron star mass  & $1.9^{+0.7}_{-0.5}$ \msun & \cite{GimenezGarcia:2016} \\
Neutron star magnetic field & $2.6\times 10^{12}$~G & \cite{Staubert:2019} \\
Pulse period       & $\sim$283~s (variable)   & \cite{Kreykenbohm:2008} \\
\hline
\end{tabular}
\end{table*}

Understanding an accreting X-ray pulsar system in detail requires studies at many different length scales. For \vx this ranges from the 10's of millions of km of the binary separation down to less than a km for the accretion column producing the X-rays. Different diagnostics across multiple wavelengths probe different elements and distance scales. 

\section{Stellar Wind}
\begin{table*}
\caption{Terminal wind velocity and stellar mass loss in the Vela X-1 system, as derived  by different authors
with different approaches -- see the references for details.}
\label{tab:vinf}
\begin{tabular}{lrrl}
\hline
Data from &  Velocity \vinf\  & Mass loss rate     & Reference\\
\hline
IUE & 1700 km/s & $\sim10^{-6}$\msun/yr &   \cite{Dupree:80} \\
IUE & 1100 km/s & --- & \cite{Prinja:90} \\
IUE &  600 km/s & $\sim10^{-6}$\msun/yr & \cite{vanLoon:2001} \\
Chandra HETGS & 1100 km/s $^1$ & $(1.5-2)\times10^{-6}$\msun/yr &\cite{Watanabe:2006} \\
IUE \& ESO FEROS &  
 700$^{+200}_{-100}$ km/s & $10^{-6.2\pm0.2}$\msun/yr
& \cite{GimenezGarcia:2016} \\ 
(Same data as above)     & $\sim 600$ km/s & $10^{(-6.07\,\ldots\,-6.19)}$\msun/yr
&\cite{Sander:2018}  \\ \hline
\multicolumn{4}{l}{$^1$ used value from \citet{Prinja:90}} \\
\end{tabular}
\end{table*}

One of the fundamental parameters to describe a HMXB is the velocity $v$ of the strong stellar wind. 
Assuming basic Bondi-Hoyle accretion, the accretion rate scales with $v^{-3}$ \citep[][and references therein]{OskFeldKre:2012}.
As Table~\ref{tab:vinf} shows, different authors have over the years derived very different terminal wind speeds \vinf, sometimes even based 
on the same data, but using different assumptions and approaches. In order to estimate  the wind velocity close to the neutron star from
\vinf, usually a ``$\beta$--law'' is assumed where $V(r) = \vinf\left(1-R_{\star}/r\right)^\beta$. 

Using a hydrodynamically consistent atmosphere model describing the wind stratification and including effects of X-ray illumination, \citet{Sander:2018} found that the velocity curve close to the mass donor may deviate significantly from a $\beta$--law, and in consequence the wind speed at the neutron star may be much lower than previously assumed. 
This, in turn, may lead to a very different flow of matter around the neutron star as El Mellah et al.\ discuss 
in these proceedings.

\section{Flux variations and absorption}
While being a persistent source, \vx shows erratic flux variations on a wide variety of timescales from days to 
minutes. If one takes X-ray monitor data from several orbital cycles and folds with the orbital period, one
arrives at a very stable \emph{mean} profile: the observed X-ray flux peaks around orbital phase 0.2 after the eclipse egress 
and then a gradual falls off towards late orbital phases \citep{Fuerst:2010,Falanga:2015}. This overall effect is
energy-dependent and mainly reflects the \emph{mean} absorption in the dense material present in the system, especially in the 
accretion and photoionization wakes \citep[][and references therein]{Grinberg:2017}. 
Still, the lightcurves of individual orbits 
may look very different especially between orbital phase 0.3 and 0.6 very different absorption measures have been determined at different times. As one example, for phase 0.5, with the neutron star between HD77581 and the observer, \citet{HaberlWhite:90} 
found absorption columns of a few times $10^{23}$, while \citet{Nagase:86} in another set of observations found an order of magnitude less. 

At shorter time scales, flares with durations of hours down to individual pulses have been reported by a variety of different authors. In other cases short ``off-states'' with fluxes below detection limits
have been observed, see, e.g., \citet{Kreykenbohm:2008} for examples of both extremes. 
\citet{Fuerst:2010} found that the pulse-averaged flux in hard X-rays can be very well described by a 
log-normal distribution. The short-term flux variations will be driven by a complex interplay between
density variations in the clumpy wind and accretion physics at the magnetosphere \citep{Martinez-Nunez:2017}, 
but the relative importance of different factors remains a point of debate.

Comparing these flux variations with models of clumpy winds can be tricky. A straightforward combination of 1D 
wind models and Bondi-Hoyle accretion leads to much larger predicted variations
than observed \citep{OskFeldKre:2012}. \citet{Ducci:2009} simulated a range of clump sizes to obtain a good match,
but require unrealistic large clump sizes. But modelling absorption just by gas clumps with clump sizes based
on current stellar wind models \emph{under-predicts} the observed absorption variations 
\citep{Grinberg:2017}.

Further information about structures in the wind, velocities and chemical composition can in principle be gained
from the study of X-ray fluorescence lines. For examples see, e.g., \citet{Sato:86}, \citet{Watanabe:2006}, 
or \citet{Grinberg:2017}.

In a recent, relatively surprising development, \vx has now also been detected significantly
at $\sim$100~$\mu$J by the Australia Telescope Compact Array with a flat radio spectrum (Van den Eijnden et al., in prep). The 
analysis of these data is still ongoing, but may yield information on the structure of the
accretion flow close to the neutron star.

Broadband spectroscopy and analysis of the pulse profile yield information about the ``last mile'', the matter in the 
accretion column close to the neutron star. \vx shows the typical X-ray spectrum for an accreting X-ray
pulsar with an absorbed powerlaw turning over at high energies, plus an iron fluorescence line.
In addition, there are two cyclotron resonance scattering features (CRSF) at $\sim 25$ and $\sim 55$\,keV 
of varying relative strength \citep{Fuerst:2014}.  The CRSF centroid energies are mildly positively correlated
with flux \citep{Fuerst:2014}, indicating accretion in the sub-critical regime. 
Swift-BAT data indicate found a long-term decrease in the centroid of the harmonic cyclotron line 
until about 2012 \citep{LaParola:2016,Ji:2019}; the physics behind this are not understood.

The pulse profile of \vx is complex with up to 5 peaks at lower energies and two asymmetric peaks in the hard X-rays 
\citep{Raubenheimer:90}. The general shape is largely stable against flux variations, but 
\citet{Doroshenko:2011} found a changed pulse pattern during an ``off-state''. Determining the emission geometry 
from these patterns is a complex task, including a full general relativity treatment. While earlier attempts have been 
undertaken in the past \citep[e.g.,][]{Bulik:95}, no recent decomposition has been done.  

\begin{figure}
\includegraphics[width=\columnwidth,clip=true]{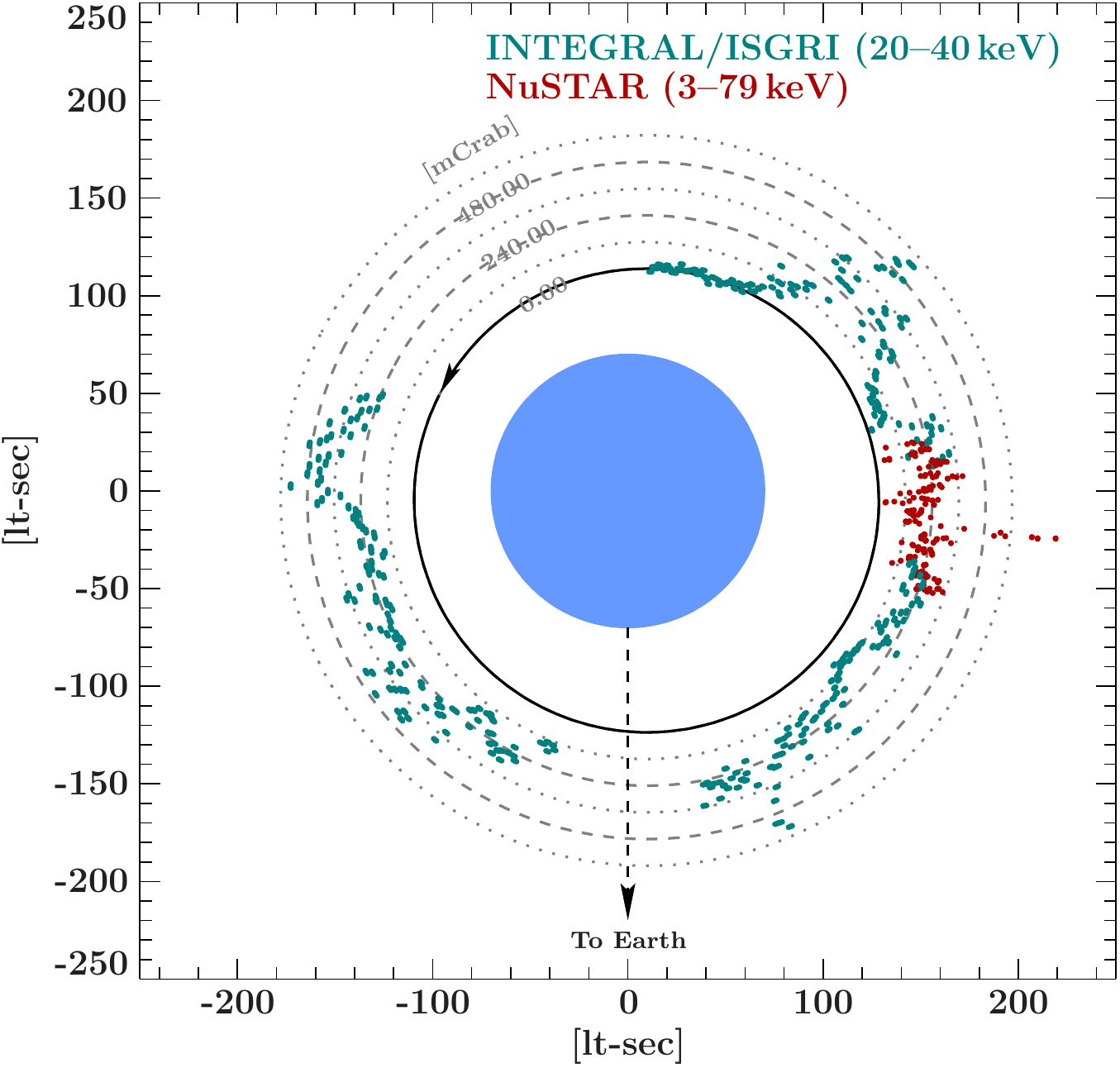}
\caption{%
INTEGRAL and NuSTAR count rates plotted over the shape of the neutron star's orbit for the observing campaign in January 2019, covering most of the binary orbit. The longer data gaps in the INTEGRAL data are due to perigee passages of the satellite.
}
\label{fig:jan2019}
\end{figure}

\section{Outlook to the future}
Despite an enormous amount of published information, many details of the \vx system are still to be determined.
We are working on various angles, e.g.: further deep studies of X-ray fluorescence lines; a deeper understanding 
of the accretion flow close to the neutron star; more realistic accretion column emission models; and multiwavelength
studies including optical and radio. A recent deep observing campaign, see Fig.~\ref{fig:jan2019}, awaits to
be analysed further in the near future.

\begin{footnotesize}
\bibliographystyle{aa} 
\bibliography{velax1_lab}
\end{footnotesize}
\end{document}